\documentclass[
  reprint,
 superscriptaddress,
 amsmath,amssymb,
 aps,
 prl,
 floatfix,
]{revtex4-1}

\usepackage{graphicx}
\usepackage{epstopdf, epsfig}
\usepackage[utf8]{inputenc}
\usepackage[T1]{fontenc}
\usepackage{amsmath}
\usepackage[usenames,dvipsnames]{color}
\usepackage{float}
\usepackage{mathtools}
\usepackage{amsmath,amssymb,amsbsy}
\usepackage{bbm,bm,mathrsfs,yfonts}
\usepackage[capitalize]{cleveref}
\usepackage{xcolor}
\usepackage{microtype}
\usepackage[export]{adjustbox}
\usepackage[normalem]{ulem} 

\newcommand\redsout{\bgroup\markoverwith{\textcolor{red}{\rule[0.5ex]{2pt}{0.4pt}}}\ULon}
\newcommand{\be}{\begin{equation}}
\newcommand{\ee}{\end{equation}}

\renewcommand{\l }{\left}
\renewcommand{\r }{\right}

\DeclarePairedDelimiter\floor{\lfloor}{\rfloor}

\begin{document}

\title{Theory of the Drift-Wave Instability at Arbitrary Collisionality}

\author{R. Jorge}\email{rogerio.jorge@epfl.ch}
    \affiliation{École Polytechnique Fédérale de Lausanne (EPFL), Swiss Plasma Center (SPC), CH-1015
Lausanne, Switzerland}
    \affiliation{Instituto de Plasmas e Fusão Nuclear, Instituto Superior Técnico, Universidade de Lisboa, 1049-001 Lisboa, Portugal}
\author{P. Ricci}
    \affiliation{École Polytechnique Fédérale de Lausanne (EPFL), Swiss Plasma Center (SPC), CH-1015
Lausanne, Switzerland}
\author{N. F. Loureiro}
    \affiliation{Plasma Science and Fusion Center, Massachusetts Institute of Technology, Cambridge, Massachusetts 02139, USA}

\begin{abstract}
A numerically efficient framework that takes into account the effect of the Coulomb collision operator at arbitrary collisionalities is introduced.
Such model is
based on the expansion of the distribution function on a Hermite-Laguerre polynomial basis, to study the effects of collisions on magnetized plasma instabilities at arbitrary mean-free path.
Focusing on the drift-wave instability, we show that our framework allows retrieving established collisional and collisionless limits. At the intermediate collisionalities relevant for present and future magnetic nuclear fusion devices, deviations with respect to collision operators used in state-of-the-art turbulence simulation codes show the need for retaining the full Coulomb operator in order to obtain both the correct instability growth rate and eigenmode spectrum, which, for example, may significantly impact quantitative predictions of transport.
The exponential convergence of the spectral representation that we propose makes the representation of the velocity space dependence, including the full collision operator, more efficient than standard finite difference methods.
\end{abstract}

\maketitle

Drift-waves (DW) are low-frequency modes that arise in a magnetized plasma when a finite pressure gradient is present, and are driven unstable when electron adiabaticity is broken, such as in the presence of finite resistivity, electron inertia or wave-particle resonances.
Due to the ubiquitous presence of pressure gradients and adiabaticity-breaking mechanisms in plasmas, the DW instability plays a role in many plasma systems \cite{Goldston1995}.
Indeed, DW are known to regulate plasma transport across the magnetic field in laboratory plasmas \cite{Horton1999,Scott2002,Burin2005,Poli2008,Schaffner2012,Mosetto2013}, and are also thought to be relevant for the understanding of fundamental transport processes occurring in active galactic nuclei \cite{Saleem2003}, dense astrophysical bodies \cite{Wu2008}, the Earth's magnetosphere \cite{Shukla1980}, and dusty plasmas \cite{Salimullah2009}.
In addition, the understanding of DW is crucial since the physics underlying a number of important plasmas instabilities, such as the electron- and ion-temperature gradient modes, resistive modes, and ballooning modes \cite{Stix1992}, relies on the same mechanisms at play in DW.

Although DW are the subject of a large number of previous studies, the effect of collisionality on the linear properties of these modes remains insufficiently understood. 
This is particularly worrisome since collisionality has been found to have both stabilizing \cite{Stix1992} and destabilizing effects \cite{White2014} on DW.
The difficulty associated with an accurate assessment of collisional effects is related to the integro-differential character of the Coulomb collision operator, $C_{ab}$, describing collisions between species $a$ and $b$ \cite{Helander2002}. Indeed, this operator cannot easily be expressed in the {guiding center} coordinate system appropriate to describe magnetized plasmas \cite{Li2011}, which often leads to its replacement by approximate, somewhat \textit{ad hoc}, operators.
As a consequence, while collisional effects on non-magnetized plasma waves, such as electron-plasma \cite{Brantov2012,Banks2017} and ion-acoustic waves \cite{Epperlein1994}, have been exhaustively characterized {at arbitrary collisionality}, such studies have not been applied to magnetized plasma instabilities, such as DW, yet.
Previous studies on the DW instability at finite collisionality have usually relied on simplified collision operators \cite{Angus2012}, or on fluid models such as the Hasegawa-Wakatani \cite{Hasegawa1983} or the drift-reduced Braginskii model \cite{Ricci2012a}, which assume that the electron and ion collision frequencies are high enough so that the particle mean free path stays small when compared with the mode parallel wavelength, $k_\parallel \lambda_{mfp} \ll 1$.

The present work overcomes this long-standing issue and provides an efficient framework, which can be easily extended to a large number of instabilities, to properly study the effect of collisionality in DW at arbitrary mean free path.
Here, we focus on the case where the DW driving mechanism is provided by the density gradient, usually referred to as the universal instability \cite{Landreman2015}, in a shearless slab geometry.
The DW growth rate that we evaluate matches both the collisionless and fluid regimes at low and high collision frequencies, respectively, and shows important deviations from the collisional limit already at $k_\parallel \lambda_{mfp} \sim 0.1$.
Furthermore, at low-to-intermediate collisionality values, the regime of interest for future tokamak devices such as ITER \cite{Aymar2002}, we show the need to retain the full Coulomb collision operator.
Indeed, the DW growth rate deviates by factors of order unity from fluid and kinetic models based on approximate collision operators such as the Lenard-Bernstein \cite{Lenard1958} and the Dougherty \cite{Dougherty1964} operators.
These are operators of the Fokker-Planck type that obey the H-theorem, contain pitch-angle scattering, have important conservation properties and, by being implemented in a number of advanced kinetic codes, are used in recent studies of DW-like turbulence, both in the core \cite{Hatch2013,Nakata2016,Grandgirard2016,Mandell2018} and at edge \cite{Shi2017,Pan2018} regions of tokamak devices.
Since quasi-linear transport models estimate the turbulence drive by evaluating the linear instability growth rate \cite{Chen2000,Bourdelle2015}, quantitative differences in the growth rate  have a large impact on the prediction of the level of transport, in particular by affecting the threshold for $\bm E \times \bm B$ shear flow stabilization.
Similarly, the linear growth rate, together with the gradient removal hypothesis \cite{Ricci2013}, is used to predict the scrape-off layer width, a parameter crucial to the overall performance of present and future tokamak devices such as ITER \cite{Halpern2013}.
Therefore, our results can impact ITER operation and the design of future fusion devices.

In addition to the instability growth rate, the framework we propose allows the evaluation of the spectrum of the linear eigenmodes.
The spectrum of collisional eigenmodes, contrary to the collisionless case, is composed of a discrete set of roots, as first shown in \cite{Ng1999}.
Deviations between the results based on the Coulomb and both the Lenard-Bernstein and Dougherty collision operators are particularly evident.
The clear differences question our current understanding of plasma turbulence.
In fact, several DW turbulence studies have shown that subdominant and stable modes can be nonlinearly excited to finite amplitude \cite{Terry2006,Hatch2011,Hatch2011a,Pueschel2016} and have a major role in nonlinear energy dissipation and turbulence saturation, affecting structure formation, as well as heat and particle transport.
The computation of such modes relies on the correct evaluation of the eigenmode spectrum. 
{As we show, this displays large changes between the Coulomb and approximate collision operators}.

Under the drift approximation (i.e., spatial scales of the fluctuations  large compared to the ion Larmor radius and frequencies smaller than the ion gyrofrequency), the framework to properly treat arbitrary collisionalities is provided by the drift-kinetic equation,
\begin{equation}
    \frac{\partial F_a}{\partial t} +  \left(v_\parallel \bm b + \bm v_{E}\right) \cdot \nabla F_a-\nabla_\parallel \phi \frac{q_a}{\sigma_a^2} \frac{\partial F_a}{\partial v_\parallel}= \sum_b \left<C_{ab}\right>
    \label{eq:dk}
\end{equation}

\noindent where $F_a=F_a(\bm R, v_\parallel, \mu, t)$ is the guiding center distribution function of the species $a$, which depends on the guiding center coordinate $\bm R$, the component of the velocity parallel to the magnetic field $v_\parallel$, the first adiabatic invariant $\mu = m_a v_\perp^2/2 B$, and time $t$ \cite{Hazeltine1968}.
The charge $q_a$, the electrostatic potential $\phi$, the parallel and perpendicular scale lengths, $v_\parallel$, and $t$, are normalized to $e$, $T_{e0}/e$, $L_n$, $\rho_s=c_s / \Omega_i$, $c_s=\sqrt{T_{e0}/m_i}$,  $c_s/L_n$ respectively, with $T_{e0}$ a reference temperature, $L_n$ the background density gradient length, and $\Omega_i = e B / m_i$. In addition,  $\bm v_E = (L_n/\rho_s) \bm b \times \nabla \phi   $ is the dimensionless $\bm E \times \bm B$ velocity, $\sigma_a = \sqrt{m_a/m_i}$, $\left<C_{ab}\right>=\int_0^{2\pi} d\theta C_{ab}/(2\pi)$ is the gyroaverage operator with $\theta$ the gyroangle, and the Coulomb collision operator is given by
\be
    \begin{split}
        C_{ab}=\frac{\nu_{ab}}{N_0}&\partial_{\bm v} \cdot \l[\frac{m_a}{m_b}({\partial_{\bm v} H_b})F_a - \partial_{\bm v}(\partial_{\bm v}G_b)\cdot \partial_{\bm v}F_a\r],
    \end{split}
    \label{eq:caa}
\ee

\noindent with $\nu_{ab}$ the characteristic collision frequency between species $a$ and $b$ normalized to $c_s/L_n$, and $H_b=2v_{tha}^3\int {F_b(\bm v')}/{|\bm v - \bm v'|}d\bm v'$  and $G_b=v_{tha}^3\int F_b(\bm v')|\bm v - \bm v'|d\bm v'$ the Rosenbluth potentials \cite{Helander2002}.
The drift-kinetic equations are coupled to the quasineutrality condition $\sum_a q_aN_{a}\left(1+\sigma_a^2{\nabla_\perp^2 \phi}{}\right)=0$ \cite{Jorge2017}, where $N_a = \int F_a dv_\parallel d \mu d\theta B/(m_a N_{0})$.

We linearize \cref{eq:dk} by expressing $F_a = F_{aM}(1 + f_{a1})$ with $f_{a1}\ll 1$ and $F_{aM}$ an isotropic Maxwellian equilibrium distribution function of constant temperature $T_{a0}$ and of density $N_{0}$ that varies perpendicularly to the magnetic field on the $L_n$ scale.
This yields

\begin{equation}
\begin{split}
    (\gamma+i k_\parallel v_\parallel)f_{a1} = i\left(k_\perp - q_a k_\parallel v_\parallel \right)\phi+\frac{\sum_b \left< C_{ab} \right>}{F_{aM}},
\end{split}
\label{eq:dklinear}
\end{equation}
\noindent where $C_{ab}$ is now the linearized version of the collision operator in \cref{eq:caa}, $\gamma$ is the growth rate, $k_\parallel$ is the wave-number parallel to $B$, and $k_\perp$ is the wave-number along the direction perpendicular to both the magnetic field and the direction of the equilibrium density gradient. 
In this work, we solve \cref{eq:dklinear} at arbitrary collisionality by expanding the distribution function into an orthogonal Hermite-Laguerre polynomial basis, i.e. $f_{a 1} = \sum_{p,j} {N_a^{pj}}H_p({v_\parallel \sigma_a/\sqrt{2 \tau_a}})L_j(\mu B/T_{a0})/{\sqrt{2^p p!}}$ with $H_p(x)=(-1)^p \exp(x^2)d^p_{x}\exp(-x^2)$ the physicists' Hermite polynomials, $L_j(x)=\exp(x)d^j_x[\exp(-x)x^j)]/j!$ the Laguerre polynomials, and $\tau_a=T_{a0}/T_{e0}$.
While a number of previous works show that the use of Hermite polynomial expansions of the parallel velocity coordinate $v_\parallel$ is advantageous \cite{Zocco2011,Loureiro2015}, both for numerical implementation \cite{Zocco2011,Bratanov2013} and to make analytical progress when the analysis of the Boltzmann equation is very complex otherwise \cite{Schekochihin2016}, the use of Laguerre polynomial expansions of the $\mu$ coordinate is recent and has only been applied in the low-collisionality regime \cite{Zocco2015,Mandell2018}.
By projecting \cref{eq:dklinear} into a Hermite-Laguerre basis, an infinite system of algebraic equations (henceforth called moment hierarchy) for the evolution of the coefficients of the expansion of $f_{a1}$, $N_{a}^{pj}$, is obtained

\begin{align}
    \gamma N_a^{pj} &= -i k_\parallel \frac{\sqrt{ \tau_a}}{\sigma_a}\left(\sqrt{{p+1}{}}N_a^{p+1 j}+\sqrt{{p}}N_a^{p-1 j}\right)\nonumber\\
    &
    +i\phi \left(k_\perp \delta_{p,0}-\frac{q_a k_\parallel}{\sqrt{\tau_a}\sigma_a}\delta_{p,1}\right) \delta_{j,0}+\sum_b C_{ab}^{pj},
\label{eq:dwmomenthierarchy}
\end{align}

\noindent with $C_{ab}^{pj}=\int \left< C_{ab} \right> H_p L_j dv_\parallel d\mu 2\pi c_s B/(N_0 m_a \sqrt{2^p p!})$ the projection of the Coulomb collision operator $C_{ab}$ onto a Hermite-Laguerre basis.

The Hermite-Laguerre moments of the linearized collision operator $C_{ab}^{pj}$ are obtained by leveraging the work
in \cite{Ji2006}, where $C_{ab}$ is projected onto a tensorial Hermite and associated Laguerre basis, $\bm p^{lk}=\bm P^{l}(\bm c)L_k^{l+1/2}(c^2)$, with the tensorial Hermite polynomials defined by the recurrence $\bm P^{l+1}(\bm c)=\bm c \bm P^l(\bm c)-c^2 \partial_{\bm c}\bm P^l(\bm c)/(2l+1)$, being $\bm P^1(\bm c)=1$ and $\bm c = \bm v/v_{th}$, and the associated Laguerre polynomials $L_k^{l+1/2}(x)=x^{-l-1/2}\exp(x)d^k_x[\exp(-x)x^{k+l+1/2}/k!]$.
This leads to
\begin{equation}
    C_{ab}=\sum_{k,l,p}\bm p^{lk} \cdot (A_{ab}^{lpk}\bm m_a^{lk}+B_{ab}^{lpk}\bm m_b^{lk}),
    \label{eq:jimoment}
\end{equation}
having expanded the distribution function as $f_{a1}=\sum_{l,k}\bm m_a^{lk} \cdot \bm p^{lk}$.
The expressions for $A_{ab}^{lpk}$ and $B_{ab}^{lpk}$ are given in Ref. \cite{Ji2006}.
In order to evaluate $C_{ab}^{pj}$, the operator $C_{ab}$ in \cref{eq:jimoment} is gyroaveraged.
This is done using the gyroaveraging identity $\int_0^{2\pi} \bm P^l(\bm c) d\theta = 2 \pi c^l P_l(v_\parallel/v)\bm P^l(\bm b)$, with $P_l(x)=d^l_x(x^2-1)^l/(2^l l!)$ the Legendre polynomials.
Then, a basis transformation from the tensorial Hermite and associated Laguerre to the Hermite-Laguerre basis is performed, $c^l P_l(v_\parallel/v)L_k^{l+1/2}(c^2)=\sum_{p=0}^{l+2k}\sum_{j=0}^{k+\floor{l/2}}T_{lk}^{pj}H_p(v_\parallel{\sigma_a}/{\sqrt{2\tau_a}})L_j({\mu B}/{T_{a0}})$, where the expressions for $T_{lk}^{pj}$ are derived in Ref. \cite{Jorge2017}.
This yields

\begin{align}
    C_{ab}^{pj}=\sum_{k=0}^{\infty}\sum_{l=0}^{p+2j}\sum_{t=0}^{j+\floor{p/2}}&\frac{(T^{-1})^{lt}_{pj} 2^l (l!)^2}{(2l)!\sqrt{2^p p! \alpha_k^l}}\frac{c_s}{N_0 (2l+1)}\nonumber\\
    &\times
    {\left(\mathcal{N}_a^{lk} A_{ab}^{ltk}+\mathcal{N}_b^{lk} B_{ab}^{ltk}\right)},
\label{eq:cabpj}
\end{align}

\noindent having introduced the normalization factor $\alpha_k^l=l!(l+k+1/2)!/(2^l(l+1/2)!k!)$,  the inverse transformation coefficients $(T^{-1})_{pj}^{lt}={\sqrt{\pi}2^p p!(l+1/2)t!}/{(t+l+1/2)!}T_{lt}^{pj}$ and the guiding center moments $\mathcal N_{a}^{lk} = \bm P^l(\bm b) \cdot \bm m^{lk} (2l)!/[2^l (l!)^2]=  \sum_{s=0}^{l+2k}\sum_{r=0}^{k+\floor{l/2}}T_{lk}^{sr}N_a^{sr} \sqrt{2^s s!/\alpha_k^l}$.
For inter-species collisions, we take advantage of the smallness of the electron to ion mass ratio $\sigma_e$ to derive simpler expressions for $A_{ab}^{ltk}$ and $B_{ab}^{ltk}$ (see, e.g., Refs. \cite{Helander2002,Ji2006}).

A closed form solution for the DW moment-hierarchy can be given in the collisionless case $C_{ab}=0$ by dividing the Boltzmann equation, \cref{eq:dklinear}, by the resonant $\gamma+i k_\parallel v_\parallel$ factor, multiplying by the Hermite-Laguerre polynomial basis functions, and integrating over velocity space, yielding
\begin{align}
    {N_a^{pj}}&=\left(-\frac{q_a \xi_a}{\tau_a}+\frac{\sigma_a k_\perp}{k_\parallel\sqrt{\tau_a}}\right)\frac{(-1)^p}{\sqrt{2^p p!}}Z^{(p)}\left(\xi_a\right) \phi \delta_{j,0}\nonumber\\
    &
    -\frac{q_a}{\tau_a} \phi\delta_{p,0}\delta_{j,0},
\label{eq:npjepw}
\end{align}
where $Z^{(p)}(\xi_a)$ is the $p$th derivative of the plasma dispersion function $Z(\xi_a)=Z^{(0)}(\xi_a)$, defined by $Z^{(p)}(\xi_a)={(-1)^p}\int_{-\infty}^{\infty}{H_p(x)e^{-x^2}}/(x-\xi_a)dx/{\sqrt{\pi}}$
\noindent and $\xi_a = \omega \sigma_a/(k_\parallel \sqrt{2\tau_a})$.
Equation (\ref{eq:npjepw}) generalizes the Hermite spectrum obtained for plasma waves \cite{Kanekar2015} and extends Hammet-Perkins-like collisionless closures obtained for $N_a^{30}$ and $N_a^{40}$ \cite{Hammett1992a} to a moment $N_a^{pj}$ of arbitrary order in a form ready to be used.

The Chapman-Enskog procedure with truncation of the moment hierarchy in \cref{eq:dwmomenthierarchy} at $p=3$ and $j=1$ can be used in the high collisionality limit, $k_\parallel \lambda_{mfp}\ll 1$.
Neglecting sound wave coupling and assuming cold ions, this yields the continuity and electron temperature equations, $\gamma N=-i( k_\parallel V - k_\perp \phi)$ and  $\gamma T=- i k_\parallel c_V V- k_\parallel^2 (\chi_{\parallel} T+ 0.12 \Delta T)/\nu$, the vorticity equation, $  k_\perp^2\gamma\phi =i k_\parallel V $, Ohm's law, $\sigma_e^2 \gamma V=i k_\parallel(\phi-N - c_T T - 0.90 \Delta T) - \nu V$, and temperature anisotropy variation $\gamma \Delta T = -12.02 \nu \Delta T/\sigma_e^2 - 2.71 i k_{\parallel} V-k_\parallel^2(0.55 T + 0.52 \Delta T)/\nu $ with $N=N_e^{00}$ the electron density normalized to $N_0$, $V= N_e^{10}/\sigma_e$ the electron parallel fluid velocity normalized to $c_s$, $T=N_e^{20}+N_e^{01}$ the electron temperature normalized to $T_{e0}$, $\Delta T = N_e^{30}-N_e^{01}$ the temperature anisotropy normalized to $T_{e0}$, $\nu$ the Spitzer resistivity normalized to $c_s/L_n$, and the numerical coefficients $(c_T,c_V,\chi_{\parallel})=(1.26,1.88,0.46)$.
When temperature anisotropy is neglected (i.e., $\Delta T=0$), the following dispersion relation is obtained:
\begin{equation}
\begin{split}
    &\sigma_e^2 \gamma^3 + \nu \gamma^2 + \frac{1+ k_\perp^2}{k_\perp^2}k_\parallel^2 \gamma-\frac{i k_\parallel^2}{k_\perp}+\frac{c_V c_T k_\parallel^2 \nu \gamma^2}{\nu \gamma+\chi_{\parallel}{k_\parallel^2}}=0,
\end{split}
\label{eq:drfluid}
\end{equation}
which reduces to the drift-reduced Braginskii dispersion relation that has similar numerical coefficients $(c_{V},c_{T},\chi_{\parallel})=(1.14,1.71,1.07)$ \cite{Zeiler1997} (we have checked that the values of the coefficients $(c_T,c_V,\chi_{\parallel})$ approach those computed by Braginskii as the order of the closure is increased).
We also note that for resistivity driven DW ($\nu>\gamma m_e/m_i$) the peak growth rate,  $\gamma \simeq 0.12$, is found at $k_\perp \simeq 1.19$ and $k_\parallel = 1.49 \sqrt{\nu}$.
If the resistivity $\nu$ in \cref{eq:drfluid} is tuned to values lower than the ones allowed by the fluid approximation ($\nu<\gamma m_e/m_i$) an electron-inertia driven DW is obtained with a peak growth rate $\gamma \simeq 0.29$ at $k_\perp \simeq 1.00$ and $k_\parallel \simeq 0.48 \sqrt{m_e/m_i}$.


\begin{figure*}
    \centering
    \adjincludegraphics
    [width=0.99\textwidth]
    {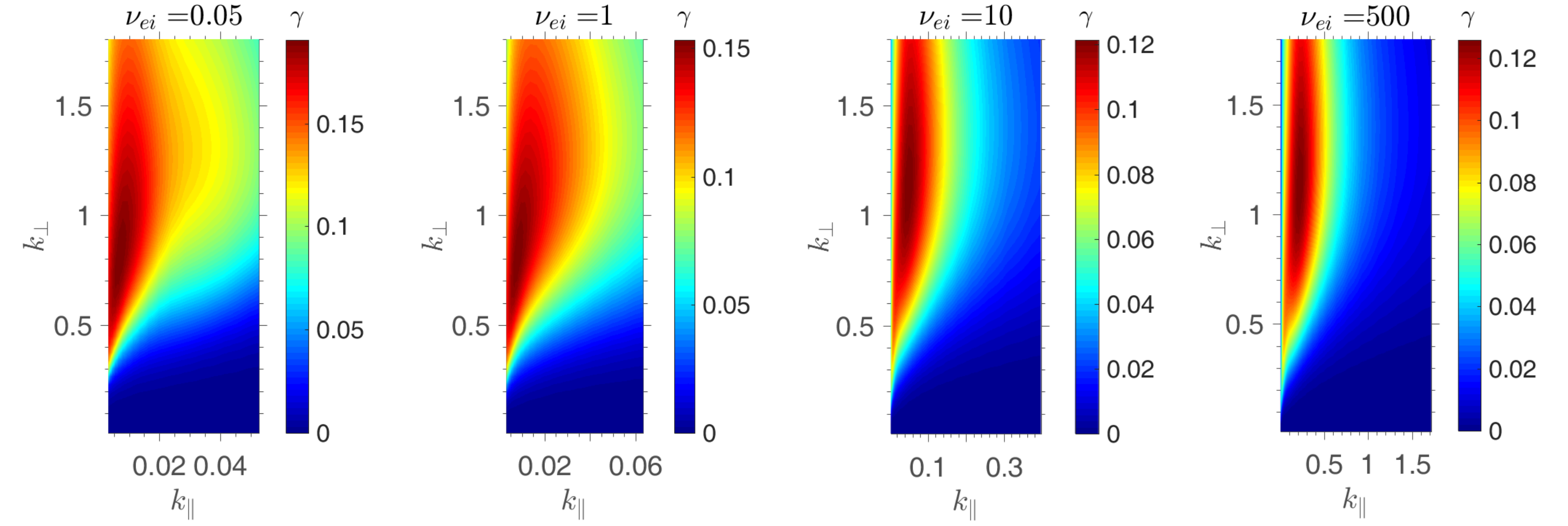}
    \caption{Growth rate of the DW instability obtained from the moment hierarchy, \cref{eq:dwmomenthierarchy}, as a function of $(k_\parallel, k_\perp)$ and, from left to right, $\nu_{ei} =0.05, 1, 10$, and $500$, in the cold-ion limit with $\sigma_e=0.023$.}
    \label{fig:dwscan}
\end{figure*}

At intermediate collisionality, the moment hierarchy equation, \cref{eq:dwmomenthierarchy}, together with Poisson equation have to be solved numerically.
In this case, a criterion to truncate the moment expansion at a suitable order $p=p_\text{max}$ and $j=j_\text{max}$ can be derived by following Ref. \cite{Schekochihin2016} where the Lenard-Bernstein operator case was considered.
This operator can be derived by setting $\partial_{\bm v} \partial_{\bm v} G_b = - \bm I v_{tha}^2/2$ and $\partial_{\bm v}H_b = \bm v$ in \cref{eq:caa}, yielding $C_{ab}^{pj} = -\nu_{ab}(p+2j)N_a^{pj}$  \cite{Lenard1958}.
The Dougherty collision operator, on the other hand, adds the necessary field-particle collisional terms to Lenard-Bernstein in order to provide momentum and energy  conservation properties, namely it sets $\partial_{\bm v} \partial_{\bm v} G_b = - \bm I T_a/m_a$, with $T_a = \int v^2 F_a dv_\parallel d\mu d\theta B/(3 N_a) $, and $\partial_{\bm v}H_b = \bm v-\bm u$, with $\bm u = \int \bm v F_a dv_\parallel d\mu d\theta B/(m_a N_a)$.
This yields $C_{ab}^{pj}=-\nu_{ab}[(p+2j)N_a^{pj}+N_a^{10}\delta_{p,1}\delta_{j,0}+T_a(\sqrt 2\delta_{p,0}\delta_{j,1}-{2}\delta_{p,2}\delta_{j,0})]$ being $T_a=(\sqrt{2}N_a^{20}-2N_a^{01})/3$ \cite{Dougherty1964,Anderson2007,Mandell2018}.
To derive the truncation criterion, we introduce the Fourier harmonics $g_{pj}=i^p \text{sgn}(k_\parallel)^p N_a^{pj}$, and insert them in the moment hierarchy equation, \cref{eq:dwmomenthierarchy}, noting that at sufficiently high index $p$, $g_p$ can be considered continuous and differentiable in $p$, and therefore $g_{p\pm 1} \simeq g_p \pm \partial_p g_p$.
By keeping only the terms proportional to $N_a^{pj}$ in the sum in \cref{eq:cabpj}, namely approximating $C_{ab}^{pj} \simeq - \nu_{ab} f_{pj} N_a^{pj}$ and effectively underestimating the collisional damping contribution of $C_{ab}^{pj}$, we obtain  $g_p\simeq g_0 \exp[-(4 \gamma \sqrt{p} + 2 \int^p f_{p j}p^{-1/2})/p_{ca}]/p^{1/4}$ at the lowest order in $1/p$, with $p_{ca}=4 |k_\parallel|\sqrt{\tau_a}/(\sigma_a \nu_{ai})$.
While for the case of the Lenard-Bernsteinand Dougherty operators, since $f_{pj} = p+2j$ for large $p$ and $j$, the solution $g_p \simeq g_0 \exp[-4(\gamma \sqrt{p}+p^{3/2}/3)/p_{ca}]/p^{1/4}$ can be obtained analytically, the coefficients $f_{pj}$ for the case of Coulomb collisions are found numerically to follow $f_{pj}\simeq A \sqrt{p}$, with $A \simeq 0.5$.
Such estimate yields
\begin{equation}
    N_a^{pj}\simeq \frac{N_0(j) i^p \text{sgn}k_\parallel^p}{p^{1/4}}\exp\left[{-\left( {\frac{p}{p_{\gamma a}}}\right)^\frac{1}{2}-2A\frac{p}{p_{c a}}}\right].
    \label{eq:coulpjspectrum}
\end{equation}
showing that the moment hierarchy can be truncated at $p_\text{max} \simeq p_{c a}$ or, if $k_\parallel \lambda_{mfp a} \gtrsim 2 \gamma^2/A$, at $p_\text{max} \simeq p_{\gamma a} = p_{c a}^2/(16\gamma^2)$ \cite{Zocco2011}.
This removes the need of \textit{ad hoc} closures for the moment hierarchy even at low collisionalities.
Regarding the truncation in $j$, since the magnetic field is uniform, no perpendicular phase-mixing in \cref{eq:dwmomenthierarchy} is present, and $j>0$ moments are present due to collisional coupling in $C_{ab}^{pj}$.
Therefore, at zero collisionality, the $j>0$ moments vanish [see \cref{eq:npjepw}].
At high collisionality, the Chapman-Enskog closure shows that $j>1$ moments are collisionally damped.
At intermediate collisionality, numerical tests show that only moments $j\le 2$ impact the growth rate.
%


\begin{figure}
 \centering
 \includegraphics[width=0.49\textwidth]{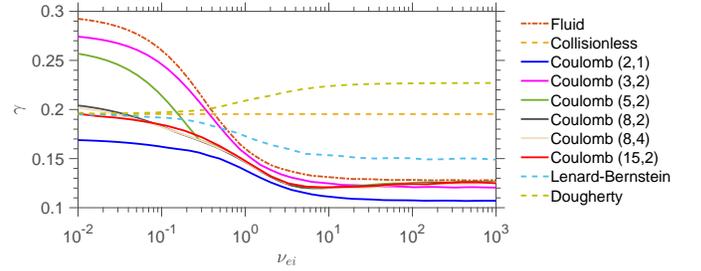}
 \caption{Comparison of the growth rate $\gamma$ maximized over $k_\parallel$ and $k_\perp$, as a function of collisionality $\nu_{ei}$, between the solution of the drift-reduced Braginskii model, the collisionless model, the linearized moment hierarchy using a simplified Lenard-Bernstein, Dougherty, and the Coulomb collision operator. A truncation at different $(p_\text{max},j_\text{max})$ is considered.}
 \label{fig:dwcoulb}
\end{figure}

The numerical solution of the moment hierarchy, \cref{eq:dwmomenthierarchy}, in the cold-ion limit with $\tau_i=0.01$ and $\sigma_e=0.023$ is shown in \cref{fig:dwscan}, where the maximum growth rate is computed over the $(k_\parallel, k_\perp, \nu_{ei})$ parameter space.
The value of $k_\parallel$ at the peak growth rate is seen to increase with $\nu_{ei}$ at large value of the resistivity, as expected from the resistive fluid dispersion relation.
For small values of resistivity it converges to $k_\parallel \simeq 0.0074 \simeq 0.32 \sigma_e$, a value close to the fluid predictions for electron-inertia driven DW.
The peak growth rate is observed to stay at $k_\perp \simeq 1$ across all values of collisionality, as also expected from the fluid theory.
By selecting the $k_\parallel$ and $k_\perp$ that yield the largest growth rate $\gamma$, \cref{fig:dwcoulb} shows a comparison between the peak growth rate resulting from the fluid model, \cref{eq:drfluid}, with the Braginskii values for  $(c_{V},c_{T},\chi_{\parallel})$, the collisionless model, \cref{eq:npjepw}, and the moment hierarchy using the Lenard-Bernstein, Dougherty, and the Coulomb collision operator solving for a different number of moments.
The linearized moment hierarchy model approaches the collisionless and the drift-reduced Braginskii model limits, at $\nu_{ei} \ll 1$ and $\nu_{ei} \gg 1$ respectively.
Deviations of the peak growth rate of the moment hierarchy from the drift-reduced Braginskii occur at values of collisionality $\nu_{ei} \lesssim 10$, and from the collisionless limit at $\nu_{ei} \gtrsim 2\times 10^{-2}$.
This corresponds to the range $0.1 \lesssim k_\parallel \lambda_{mfp} \lesssim 100$ (at the $k_\parallel$ of the peak growth rate), a range that overlaps with the regime of operation relevant for present and future tokamak machines \cite{Pitts2011}.
Deviations of up to $50\%$ with respect to the Lenard-Bernstein and Dougherty operators arise on both the peak growth rate and its corresponding $k_\parallel$ and $k_\perp$.
We note that convergence is observed for $p_{\text{max}} = 15$ and $j_\text{max}=2$ up until $\nu_{ei} \sim 10^{-1}$.
The observed value of $p_{\text{max}}$ is close to the estimate in \cref{eq:coulpjspectrum}, which for $\nu_{ei} = 10^{-1}$ and $k_\parallel \simeq 0.32 \sigma_e$ yields $p_{\text{max}} \simeq p_{ce} \simeq 13$.
We remark that pseudospectral decompositions converge exponentially with the number of modes used.
Therefore, with respect to finite-difference methods that display algebraic convergence, the framework proposed here is particularly efficient for numerical implementation.


\begin{figure}
    \centering
    \includegraphics[width=0.48\textwidth]{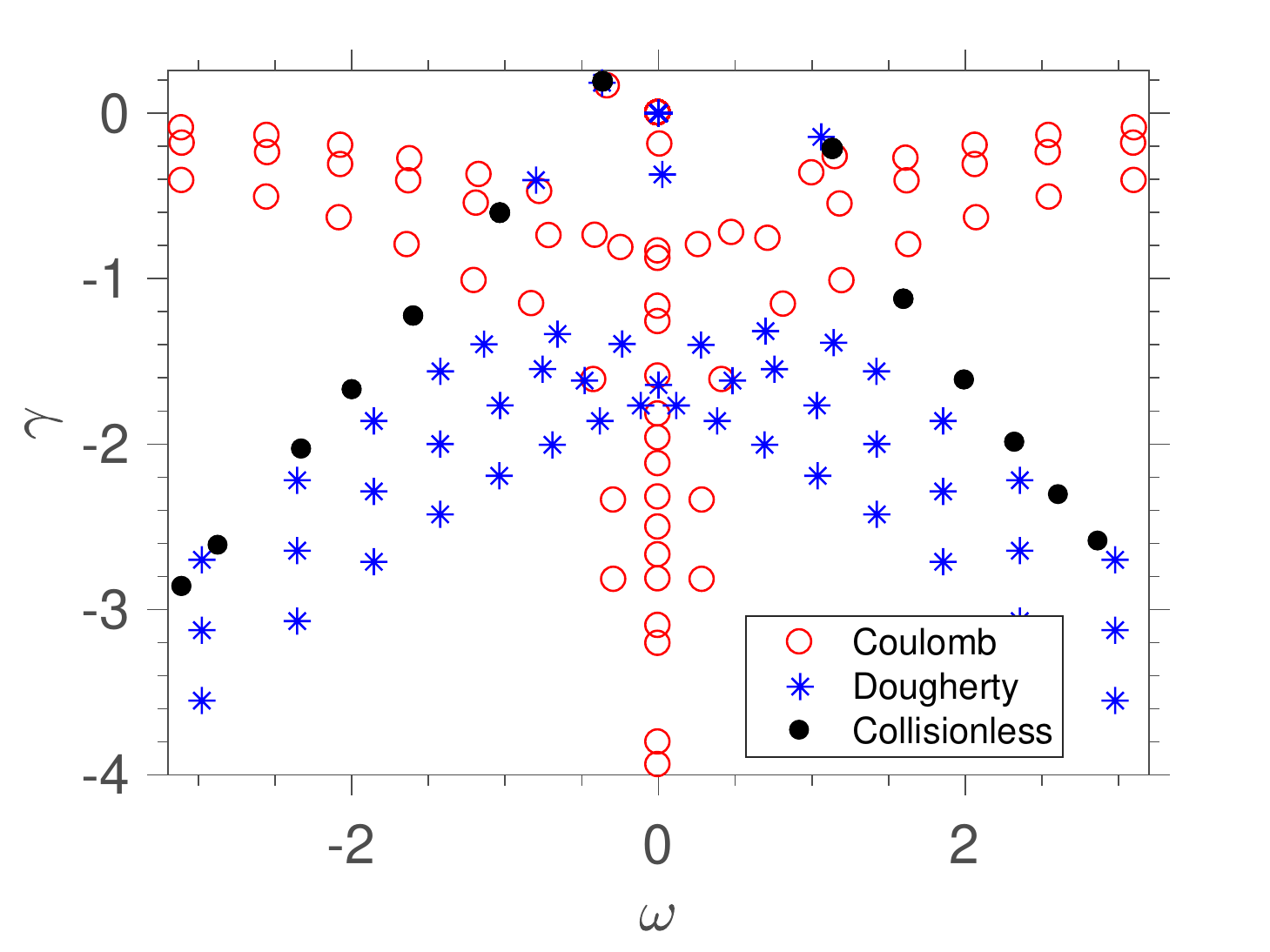}
    \caption{Eigenvalue spectra obtained with the collisionless model, the linearized moment hierarchy equation using the Coulomb and the Dougherty collision operator {at the wave-number ($k_\parallel$, $k_\perp$) corresponding to the fastest growing mode in the cold-ion limit for $\nu_{ei}=0.4$ and $\sigma_e=0.023$, with 15 Hermite and 2 Laguerre polynomials}. The analysis is carried out with $p_\text{max}=15$ and $j_\text{max}=2$.}
    \label{fig:spectra}
\end{figure}

We compare in \cref{fig:spectra} the spectra obtained with the collisionless model, and with the Dougherty and the Coulomb collision operators in the moment hierarchy at $\nu_{ei}=0.4$ for the values of $(k_\parallel, k_\perp)$ that yield the largest $\gamma$.
Figure \ref{fig:spectra} shows a clear difference between the eigenmode spectra of the two operators.
While modes with finite frequency are related to the damping of electron distribution function, modes at $\omega \ll 1$ are due to strong collisional damping of the cold-ion distribution function.
The damping rate of the electron modes decreases with the frequency when the Coulomb collision operator is considered, contrary to the Dougherty case. 
This is possibly related to the fact that the collisional drag force decreases with the particle velocity in the Coulomb collision operator and increases in the Dougherty one.
We note that the eigenmode spectrum using the Dougherty collision operator in \cref{fig:spectra} is similar to the one obtained in Ref. \cite{Bratanov2013} using a Lenard-Bernstein one.

In this Letter, for the first time, Coulomb collisions are taken into account in the description of magnetized plasma instabilities at arbitrary collisionalities, focusing on the linear properties of the DW instability.
The analysis we perform in a relatively simple configuration shows that the corrections introduced by the full Coulomb collision operator with respect to simplified collision operators, presently used in state-of-the-art codes, are qualitatively and quantitatively significant at the relevant collisionality regime of operation of future nuclear fusion devices such as ITER.
Our work provides a particularly efficient numerical framework to treat Coulomb collisions that can easily be extended to nonlinear simulations and be used to study other instabilities in magnetized plasmas.
By projecting onto a Hermite-Laguerre basis the drifts that arise in the Boltzmann equation from possible inhomogeneities of the magnetic field, instabilities such as the ballooning mode, can be described within the framework presented here.
A gyrokinetic extension of the framework introduced here will be the subject of a future work.


%
We thank S. Brunner for helpful discussions and the referees for suggestions that have improved this Letter.
This work has been carried out within the framework of the EUROfusion Consortium and has received funding from the Euratom research and training programme 2014-2018 under grant agreement No 633053, and from Portuguese FCT - Fundação para a Ciência e Tecnologia, under grant PD/BD/105979/2014, carried out as part of the training in the framework of the Advanced Program in Plasma Science and Engineering (APPLAuSE, sponsored by FCT under grant No. PD/00505/2012) at Instituto Superior Técnico (IST). 
This work was supported in part by the Swiss National Science Foundation.
N.F.L. was partially funded by US Department of Energy Grant no. DE-FG02-91ER54109.
The views and opinions expressed herein do not necessarily reflect those of the European Commission.

\bibliographystyle{unsrt}
\bibliography{library}

\end{document}